\documentclass[]{IAC_style}
\usepackage{fix-cm}

\begin{document}

\IACpaperyear{2025} 
\IACpapernumber{IAC-25,C1,IP,18,x95380} 
\IAClocation{Sydney, Australia} 
\IACdate{29 September - 3 October 2025} 

\title{Open-Source High-Fidelity Orbit Estimation for Planetary Science and Space Situational Awareness Using the Tudat Software}

\IACauthor{L. Gisolfi}{1}{1}
\IACauthor{D. Dirkx}{1}{0} 
\IACauthor{S. Fayolle}{3}{0} 
\IACauthor{V. Filice}{1}{0} 
\IACauthor{R. Alkahal}{1}{0} 
\IACauthor{M. Avillez}{2}{0} 
\IACauthor{T. Dijkstra}{1}{0} 
\IACauthor{J. Hener}{1}{0} 
\IACauthor{L. Hinüber}{1}{0} 
\IACauthor{M. Langbroek}{1}{0} 
\IACauthor{N. Maistri}{1}{0} 
\IACauthor{M. K. Plumaris}{3}{0} 
\IACauthor{A. Sanchez Rodriguez}{1}{0} 
\IACauthor{G. Cimò}{7}{0} 
\IACauthor{K. Cowan}{1}{0} 
\IACauthor{F. Dahmani}{5}{0} 
\IACauthor{J. Encarnacao}{1}{0} 
\IACauthor{G. Garrett}{1}{0} 
\IACauthor{S. Gehly}{1}{0} 
\IACauthor{X. Hu}{6}{0} 
\IACauthor{M. Jeanjean}{1}{0} 
\IACauthor{A. López Rivera}{4}{0} 
\IACauthor{A. Minervino Amodio}{1}{0} 
\IACauthor{G. Molera Calves}{8}{0} 
\IACauthor{M. Reichel}{6}{0} 
\IACauthor{B. Root}{1}{0} 
\IACauthor{M. Søndergaard}{1}{0} 
\IACauthor{D. Stiller}{10}{0} 
\IACauthor{S. Van Hulle}{1}{0} 
\IACauthor{G. Verdoes Kleijn}{9}{0} 
\IACauthor{O. R. Williams}{9}{0} 
\IACauthor{D. Witte}{1}{0}

\IACauthoraffiliation{Delft University of Technology, Kluyverweg 1, 2629HS Delft, The Netherlands}
\IACauthoraffiliation{Purdue University, West Lafayette, Indiana}
\IACauthoraffiliation{European Space Agency (ESA/ESTEC), The Netherlands}
\IACauthoraffiliation{The Exploration Company, Behringstr. 6 Planegg, BY 82152, Germany}
\IACauthoraffiliation{GMV Portugal, Alameda dos Oceanos 115, 1990-392 Lisboa, Portugal}
\IACauthoraffiliation{Universität der Bundeswehr München, Werner-Heisenberg-Weg 39, 85579 Neubiberg, Germany}
\IACauthoraffiliation{Joint Institute for VLBI ERIC, PO Box 2, 7990 AA Dwingeloo, The Netherlands}
\IACauthoraffiliation{University of Tasmania, Australia}
\IACauthoraffiliation{Rijksuniversiteit Groningen, Broerstraat 5, 9712 CP Groningen}
\IACauthoraffiliation{University of Washington, Seattle, Washington, United States}

\abstract
{The TU Delft Astrodynamics Toolbox (Tudat) is a free open-source software (FOSS) suite geared towards research and education in computational astrodynamics. It has been applied primarily to numerical
                simulation of the dynamics of objects in space, ranging from optimization of re-entry vehicle dynamics
                to the modeling of planetary spacecraft tracking, and the dynamics of natural solar system bodies. The
                powerful and versatile estimation module of Tudat has been used for a broad range of studies for both
                current and future space missions. It has the capability to combine optical and radiometric tracking data
                from multiple spacecraft with Earth-based observations into a comprehensive estimation of the dynamics
                of both natural and artificial solar system bodies, as well as physical parameters of interest. Building
                upon this general and adaptable framework, recent developments have focused on incorporating the necessary functionality required for real tracking data analysis. In this paper, we present the integration of
                these capabilities into Tudat’s fully open-source framework, with a combined focus on planetary missions
                and Space Situational Awareness (SSA). At present, the software provides capabilities to process several
                categories of observational data: (i) deep-space Doppler and range tracking data of planetary missions
                collected by the Deep Space Network (DSN) and ESA’s ESTRACK, supporting multiple formats such
                as IFMS, ODF, and TNF; (ii) deep-space Doppler and VLBI tracking data of planetary missions collected by the Planetary and Radio Interferometry and Doppler Experiment (PRIDE) with radio
                (astronomy) telescopes; (iii) optical astrometry and radar tracking archived by the Minor Planet Center
                (MPC) and the Natural Satellite Data Center (NSDC). By computing observation residuals using existing
                orbital solutions as references, we show that our observation models are accurate to the intrinsic quality of
                the data (e.g., better than 0.05 mm/s for typical deep-space Doppler data). Additionally, we demonstrate
                that our dynamical models possess the level of fidelity necessary to enable precise orbit estimation, effectively leveraging the high quality of the available tracking data. Tudat is unique in providing modular
                and flexible open-source high-fidelity modeling across a broad range of orbital regimes, enabling interdisciplinary applications. We provide an overview of the data processing and estimation capabilities, and give
                examples from various mission domains. These include high-precision orbit estimation using deep-space
                Doppler tracking data, orbit determination of cis-lunar/xGEO space debris in highly non-linear regimes (specifically targeting upper stages of lunar missions) from astrometric data, and estimation of small solar
                system bodies using astrometric data. 

}

\maketitle

\thispagestyle{fancy}


\section{Introduction}
\label{sec:introduction}

The TU Delft Astrodynamics Toolbox (Tudat) grew out of a need for a shared software framework to perform astrodynamics research and education at the Astrodynamics and Space Missions Section of the Aerospace Engineerng faculty of TU Delft.
Prior to the development of Tudat, computational astrodynamics at TU Delft relied largely on isolated scripts and tools written by individual students and researchers, or on external tools developed by research collaborators. For students and Ph.D candidates, this often involved rewriting common components - such as numerical integrators, force models, or state estimators - from scratch, leading to potentially inconsistent results, duplicated effort, and software that was difficult to maintain or share. While some internal efforts were made to consolidate tools, the absence of documentation standards and coding guidelines resulted in fragmented and unscalable codebases. With the expansion of the group in the early 2010s, Tudat started to be developed as a C++ toolbox and extensively maintained and utilized within the group, and it quickly turned into a shared, reusable, and robust software platform to support spacecraft dynamics simulation and estimation \citep{Kumar2012,Dirkx2022}. In 2020, the group began to develop a Python interface for Tudat -named Tudatpy- to improve usability and accessibility. This development drastically simplified the installation process, which now happens through the use of a \texttt{conda} package, and broadened the user base. Since then, continuous efforts have been dedicated to migrating all functionalities to Python, enhancing documentation, and improving the flexibility and usability of the codebase. 
In 2022, with a small expansion of the Tudat development team, efforts were started to leverage the strong framework for simulations to also ingest and process real tracking data. The present paper provides an overview of the current status of these developments. For more details, we encourage the reader to visit our website\footnote{https://docs.tudat.space/} that contains installation instructions, documentation and examples.

While originally developed mainly for orbit simulations, work in the European ESPaCE project \citep{Thuillot2013} led to the first developments for simulated state and parameter estimation. 
Up until 2022, many developments were made in the code to allow it to be used as a modular and flexible tool to perform such analyses for a wide range of bodies, data types and scenarios. During this period, it was used for numerous simulation studies (see below) demonstrating its versatility and maturity in propagation and (simulated) estimation. Its first applications using real tracking data was to perform orbit determination of the Lunar Reconnaissance Orbiter (LRO) using one-way laser ranging data from the International Laser Ranging Service (ILRS) stations, estimating spacecraft and clock parameters \cite{BAUER2016,Bauer2017}. In the same research project, it was used for simulated analysis of interplanetary laser ranging data \citep{Dirkx2014,Dirkx2014b,Dirkx2015,Dirkx2016,Dirkx2019b}. 

As of 2016, Tudat has been used for a range of analyses for the JUICE mission, including simulated analyses of moon dynamics and estimation strategies \citep{Dirkx2016b,FAYOLLE2022} and analyses for the contributions of various different data types \citep{Dirkx2017,Fayolle2021,Villamil2021,Fayolle2023,Fayolle2024,Hener2025,Zenk2025} to the estimation of Galilean moon gravity fields, ephemerides, tidal parameters, etc. Beyond this, it has been used for estimation-related simulations for missions that are in flight \citep{Zwaard2022,Hu2023,Fayolle2023b,Amodio2025}, in preparation \citep{Sun2024,Afzal2025,Afzal2025b}, and more speculative missions to analyze their potential science return \citep{Plumaris2022,Plumaris2024}, as well as for the development and analysis of dynamical modelling strategies for natural bodies in the context of state estimation \citep{Kumar2015,Dirkx2019,Martinez2025,Fayolle2025}. Although a lesser focus of Tudat up until recently, it has also been subject to various studies in space situational awareness, including uncertainty propagation and re-entry predictions \citep{Ronse2014,Hoogendoorn2018}, debris orbital evolution \citep{Pons2019,SCHILD2023} and orbit estimation from astrometric data \citep{Witte2024}.  A full list of publications that were made possible by the Tudat software can be found on our \href{https://docs.tudat.space/en/latest/about/research-output.html}{research output page}.

Expanding on these advances and applications in simulated analyses, Tudat has grown beyond its initial role as a simulation tool, offering robust data ingestion, processing, and analysis capabilities that effectively support real-world astrodynamics applications. Its robust estimation framework integrates diverse data types - optical astrometry, radiometric range and Doppler, VLBI - into consistent, high-fidelity orbital solutions. This paper showcases Tudat's functionalities for data ingestion, observation processing, and parameters estimation for both natural and artificial bodies. An overview of the data types currently supported, along with the available downloaders and interfaces for data retrieval is presented in \autoref{sec:DataRetrievalandIngestion}. 
To improve clarity in our analysis, we will refer to \textit{residuals} as the difference between Tudat-simulated observables generated using a reference orbit (e.g., SPICE) and actual measurements obtained from sources such as DSN, ESTRACK, or Horizons. The terms \textit{pre-fit residuals} and \textit{post-fit residuals} denote the difference between simulated observables generated from a Tudat-propagated trajectory and the corresponding observations, evaluated before and after the estimation process, respectively. 

In \autoref{sec:radiometric_observables}  we show the residuals obtained from both closed-loop Doppler observables (e.g., from DSN, ESTRACK), as well as open-loop Doppler residuals obtained using the Planetary Radio Interferometry and Doppler Experiment (PRIDE) \citep{MoleraCalves2021,Gurvits2023}. In addition, we present residuals computed using astrometric data from the Minor Planet Center (MPC) \citep{mpc_web} in \autoref{sec:data_processing:astrometry}. Tudat’s orbit estimation capabilities, featuring orbit estimation results, showing the pre/post-fit residuals for the Gravity Recovery and Interior Laboratory (GRAIL) and the Mars Reconnaissance Orbiter (MRO), as well as small solar system body ephemeris estimation and predictions for the recent Kosmos 482 descent craft reentry - are showcased in \autoref{sec:EstimationCapabilities}. Future planning, development and research projects are discussed in \autoref{sec:future_developments}. 

Aligned with the principles of open science and open-source development, the examples presented in this paper are publicly accessible via GitHub through the \href{https://github.com/tudat-team/tudatpy-examples/tree/IAC25}{\texttt{tudatpy-examples}} repository in the dedicated \texttt{IAC25} branch. It is important to note, however, that Tudat's codebase is under continuous development, and expansions to the functionalities introduced here may occur over time. 

\section{Data Retrieval and Ingestion}
\label{sec:DataRetrievalandIngestion}
\begin{figure}[!htbp]
        \includegraphics[width=1\columnwidth]{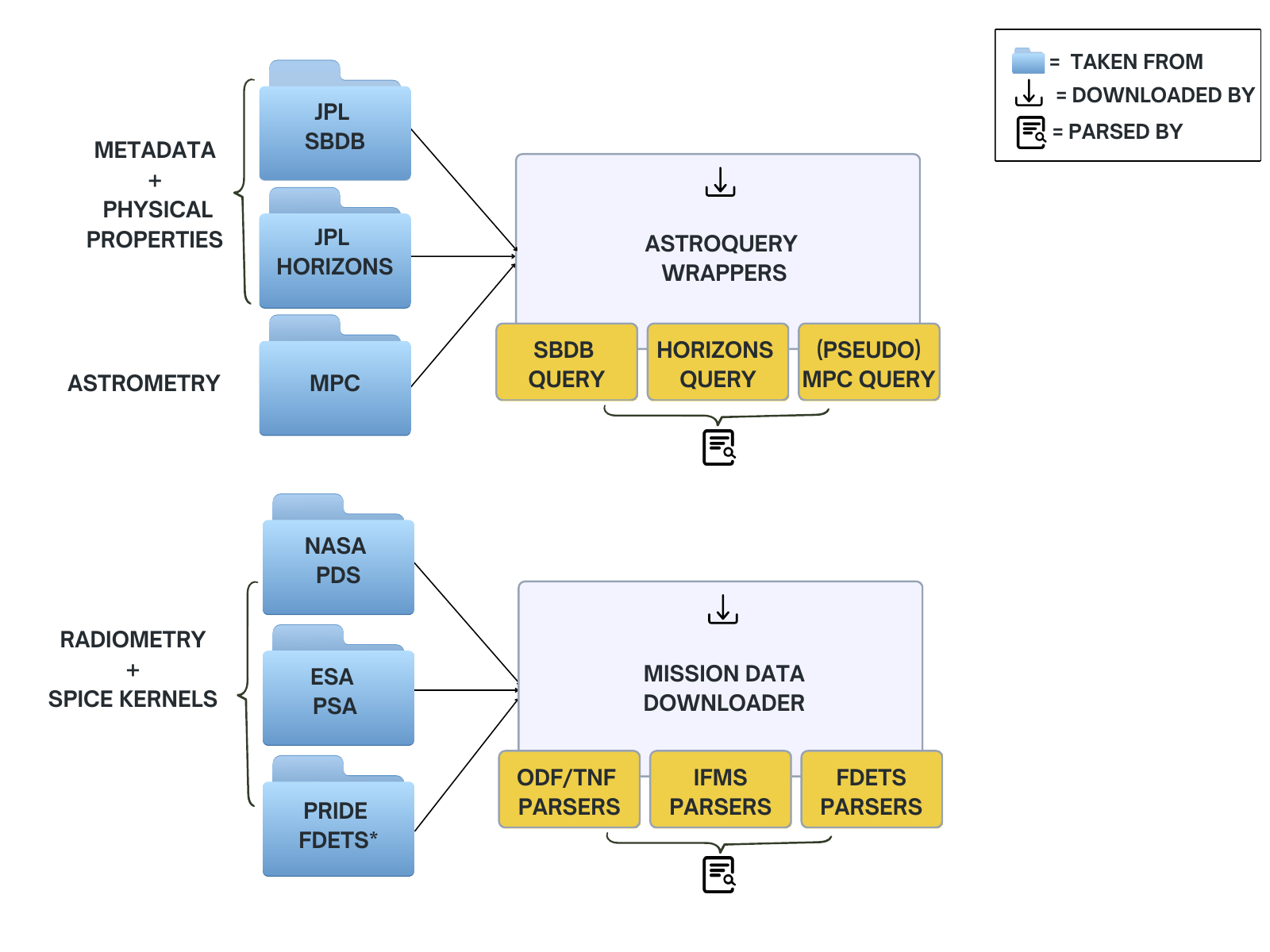}
        \caption{Visual overview of the available Tudat's interfaces, downloaders, and parsers, allowing for data and metadata retrieval and processing. 
        \small{*Although a unified server for PRIDE data is still being established, the mission data downloader will also support downloading of the Fdets files.}}
        \label{downloaders_and_interfaces}
\end{figure}

Radiometric and astrometric data, as well as other relevant quantities used throughout a typical orbit estimation routine (body masses, ephemerides, orientations, etc.), are retrieved and ingested within Tudat as shown in Figure \ref{downloaders_and_interfaces}.

DSN and ESTRACK radio science data are available through the \href{https://pds.nasa.gov/}{Planetary Data System (PDS)} and on the \href{https://psa.esa.int/psa/#/pages/home}{Planetary Science Archive (PSA)}, respectively. These public archives provide mission data in standardized bundles, organized by date intervals. Each bundle contains the final data products ready for use in orbit estimation routines and the calibration files necessary for processing the data into final products, along with relevant ancillary files, such as ASCII files linking to SPICE kernels and log files detailing the data processing.
Currently, there is no equivalent version of PDS and PSA that includes PRIDE Doppler tracking data.
However, collaboration between TU Delft, JIVE, the University of Tasmania (UTAS), and the PRIDE team is ongoing to set up such an archive \citep{PallichadathPrideDataAnalysis}. 

Optical data are taken from the MPC, and the current configuration also enables users to effortlessly convert pseudo-MPC datasets, such as the one provided by \href{https://www.projectpluto.com/}{Project Pluto}, into Tudat-compatible objects, upon manual download. Preliminary analyses have also been performed using manually ingested optical data from the Natural Satellite Data Center (NSDC) \citep{Arlot2009}. Orbital elements and ephemerides, along with other possible interesting quantities, such as body masses and predicted/reconstructed angular coordinates (i.e. simulated right ascension and declination), are retrieved from JPL Horizons \citep{Giorgini2015}, SBDB \footnote{https://ssd.jpl.nasa.gov/tools/sbdb\_lookup.html} and SPICE \cite{ACTON1996}.

The following subsections outline how Tudat retrieves and integrates the supported data types. This process primarily involves interfacing with existing astronomical tools (like \href{https://astroquery.readthedocs.io/en/latest/index.html}{Astropy's Astroquery}, see \cite{Ginsburg2019}), or utilizing specialized downloaders and parsers developed by our team.\\

\subsection{Mission Data Downloader}
At present, Tudat supports the loading of ODF \cite{trk218}, TNF \cite{trk234} and IFMS \cite{esaifms} data files for planetary radio tracking data, providing the standard information used within radio science applications. These are typically retrieved from either the PDS and the PSA through a dedicated downloader per mission, see Figure \ref{downloaders_and_interfaces}. This functionality within Tudatpy offers a streamlined approach to the downloading of radio tracking data sets, as well as the required ancillary data. For the ODF and IFMS files, we have implemented a dedicated file reader and parser to read out the contents of the files and convert these to Tudat-compatible data structures. For the TNF files, we make use of the pytrk234\footnote{\url{https://github.com/NASA-PDS/PyTrk234}} Python library to read the TNF binary file, after which we convert it into Tudat-compatible data structures.

In addition to the downloading of the data files, our downloader also enables the download of SPICE kernels, including clock, frame, orientation and planetary kernels. If desired, users can specify the mission-related meta-kernel containing the list of kernels to be downloaded. The library provides automatic downloading of Earth Orientation Parameters (EOP) data, which describe irregularities in the rotation of planet Earth and are essential to define the Earth-fixed frame with high accuracy. The library also facilitates access to ancillary ionospheric and tropospheric data that are provided by the DSN. 
Finally, it provides automatic downloading of meteorological data files in several formats, which contain information on humidity, water vapor pressure and temperature at the ground station, which can be used for (re)computation of media corrections.
Recent developments in Tudat include the integration of high-fidelity ionospheric and tropospheric corrections using global IONEX maps \citep{Schaer1998} and VMF3 gridded troposphere models \citep{Landskron2018}, which not only improve the accuracy of radiometric models, but also enable independent validation of the standard atmospheric corrections typically applied in DSN and ESTRACK data products.  The library allows to download the files needed to perform such analysis.  

At the time of writing, the library supports data downloading for the following missions: MRO, JUICE, Cassini, GRAIL, MEX and Rosetta; nevertheless, its modularity allows users to define downloaders for data of other missions, making it a flexible tool. Tudat's downloader supports customizable output paths, allowing users to control where downloaded files are stored. Future development aims to extend compatibility to additional missions such as Mars Insight, Venus Express (VEX), and Lunar Reconnaissance Orbiter (LRO).

\subsection{MPC, SBDB, Horizons and  SPICE Interfaces}
\label{sec:data-ingestion:mpc-sbdb-horizons}
Retrieving data from the MPC, JPL Horizons and JPL SBDB is managed through dedicated tudatpy modules.
Each module is developed around the corresponding query module of \texttt{astroquery} \cite{Ginsburg2019}, which performs the data requests to the APIs of the respective services.
The functionality implemented in \texttt{tudatpy} builds on top of the \texttt{astroquery} classes, adding various utilities in addition to the pre-processing required to convert the raw data into the data structures that can be integrated with Tudat. 

Astrometric observations published by the MPC are retrieved using the \texttt{astroquery} package \citep{Ginsburg2019} and preprocessed by converting the units and time scales of the observations to Tudat compatible objects.
Additionally, the observations may be filtered by observed bands, referenced star catalog, observation type, observatory and their epoch to facilitate further analysis.
Observations referencing old star catalogs can be automatically debiased following the methodology described in \cite{Eggl2020}.
This allows to correct old observations using the data from the Gaia DR2 release.
The observations are then converted to Tudat data structures, which make the observations available to the estimation pipeline.
At present, the readily available conversion only supports optical observations from ground-based or satellite-based observatories (radar data is planned to be implemented in the future).
The corresponding weights of the observations can either be set manually by the user, or using the methodology described in \cite{Veres2017}.

Our interface to the JPL Horizons system can retrieve ephemerides in the format of observer tables or Cartesian state vectors.
Using \texttt{astroquery}, the request parameters are converted from the Tudat conventions to the format required by the Horizons API.
Additionally, since queries to the JPL Horizons API are restricted in their maximum size, queries larger than the maximum permitted size can be automatically split in multiple subqueries, which are combined in the post-processing of the request for a user-friendly experience.
Similarly, a batch interface is provided, which allows to retrieve ephemeris for multiple bodies in a single query.
The retrieved data can either be output to the user, or added to the simulation environment as ephemeris settings, making it readily available to the numerical simulation. 

Properties of small solar system bodies can be retrieved from the JPL SBDB API, again using the \texttt{astroquery} package.
The SBDB API provides extensive information about the small body, including its orbit, classification and family, as well as physical parameters such as rotation parameters, spectral information, and the diameter if available.
Additionally, the SPK ID is retrieved, which is the unique ID used by the SPICE toolkit to reference the body.

Combining the three data sources allows for modular simulations with various applications in both space situational awareness and planetary science contexts, as is outlined in \autoref{sec:Astrometric_Estimation}. 

\section{Tudat's Processing Capabilities}
\label{PreFit_Residuals}
In this section, we present and validate Tudat's observation models for the radiometric and astrometric data, retrieved as described in \autoref{sec:DataRetrievalandIngestion}. The validation is performed by computing the difference between the observed and the computed ones, as predicted by Tudat's observation models. In order to isolate the performance of the observation models from potential deficiencies in the dynamical ones, we use high-precision trajectories as reference orbits. This provides a direct validation of the observation model. Specifically, for radiometric observables, the reference orbits are the reconstructed spacecraft trajectories provided by the navigation teams as SPICE kernels; for astrometry, we use the body's ephemeris as provided by JPL Horizons.  We expected the resulting residuals should exhibit the characteristics of white noise (zero mean, Gaussian distribution), particularly in cases where the reference trajectory was produced using the same tracking data. 

A critical aspect of real data analysis is the precise handling of time and timescale conversions, which Tudat performs using the Standards of Fundamental Astronomy (SOFA) library that implements the official IAU conventions (see \cite{SOFA}). Equally fundamental is the accurate modeling of Earth-based tracking station positions. To this end, Tudat implements the high-precision ITRS rotation model and ground station position variations, according to the IERS 2010 convention (see \cite{IERS2010}). The pole tide and ocean tide loading are also included, using the eleven leading harmonics. This is sufficiently accurate for planetary tracking applications, see \cite{Moyer2005}.


\subsection{Radiometric Observables}
\label{sec:radiometric_observables}
A fundamental principle underlying radiometric observables is the precise light time computation.  By modeling instrumental, relativistic, and media propagation effects as time delays added to the purely Newtonian light time, Tudat can accurately compute one-way, two-way (round-trip) and n-way light times (\cite{BERTONE2018}, \cite{Moyer2005}, \cite{Saastamoinen1972}, \cite{chao1974tropospheric}). The modular design of the software also allows for the straightforward implementation of additional user-defined corrections. 
The following subsections detail the validation of Tudat's observation models for closed-loop, sequential range, and open-loop observables.

\subsubsection{Closed-Loop Doppler and Sequential Range}
\label{subsec:closed_loop_prefit}
Tudat allows to model both two and three-way sequential range and two and three-way closed-loop Doppler observables acquired by DSN and ESTRACK, with the detailed observation models defined by \cite{Moyer2005}. The uplink frequency for these observables can be modeled as either constant or ramped, with the ramp coefficients being read directly from the tracking data files. 
The analysis presented here validates both the averaged Doppler and sequential range observation models on two representative cases: the GRAIL mission for operations in the Earth system (using ODF files),  and the MRO for an interplanetary mission (using TNF files). We show Doppler residuals (using the missions' Spice kernels to compute their states) for four different passes during April 2012 for GRAIL, and both Doppler and sequential range residuals for the full year of 2012 for MRO. The Doppler data provided in the tracking files typically have a short integration time, such as 1 second. However, we compress the data and present our results at a more typical integration time of 60 seconds. MRO uses an X-band radio link for radiometric tracking, while GRAIL uses an S-band system.

\begin{figure*}[h]
    \centering
    \begin{subfigure}[b]{0.83\textwidth}
        \includegraphics[width=0.83\textwidth]{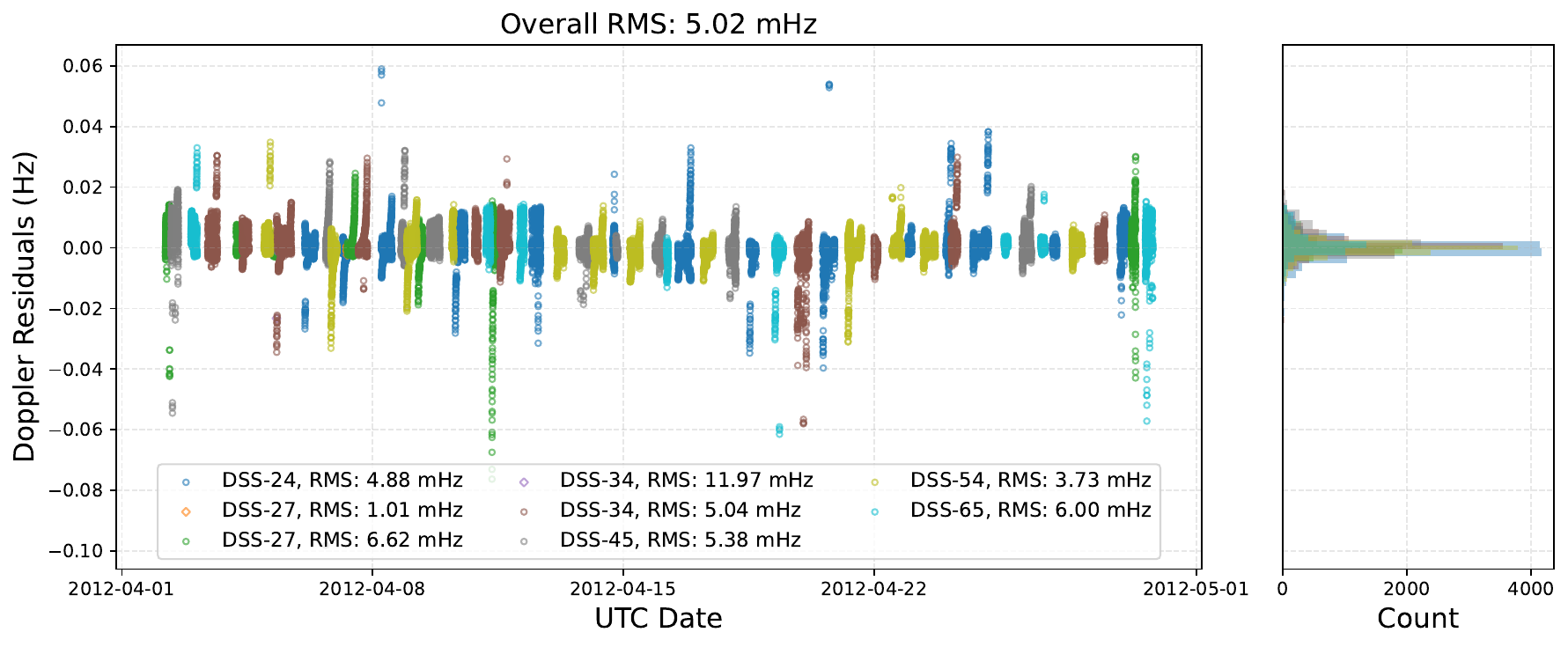}
        \caption{GRAIL two-way averaged residuals over April 2012. Different link end receivers have different colors.}
        \label{fig:grailDopplerSpice}
    \end{subfigure}
    \vfill
    \begin{subfigure}[b]{0.83\textwidth}
        \includegraphics[width=0.83\textwidth]{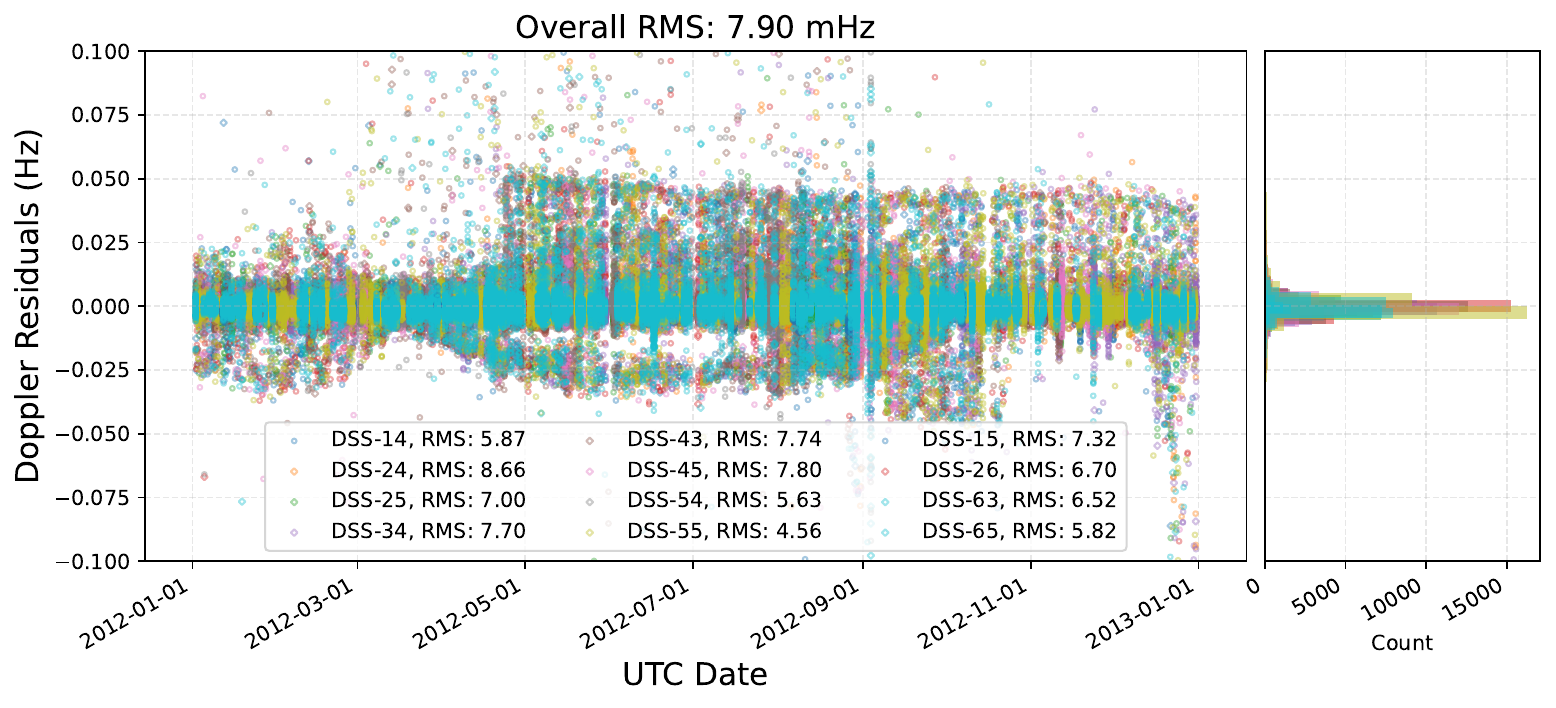}
        \caption{MRO averaged Doppler residuals over the year 2012. Diamonds indicate three-way Doppler; circles indicate two-way Doppler. Different link end receivers have different colors. RMS values are expressed in millihertz.}
        \label{fig:mroDopplerSpice}
    \end{subfigure}
    \caption{GRAIL (Panel a) and MRO (Panel b) averaged Doppler residuals.}
    \label{fig:DopplerSpice}
\end{figure*}

\autoref{fig:DopplerSpice} shows the Doppler residuals for both missions, where outliers greater than 0.1 Hz have been removed.  
For both cases, modeling the correct antenna position has proven to be crucial to obtaining flat residuals. The GRAIL Doppler residuals (\autoref{fig:grailDopplerSpice}, with an RMS of $0.005 \ Hz$, show a minor, unresolved systematic signature in some tracking passes. The MRO Doppler residuals (\autoref{fig:mroDopplerSpice}), with an RMS of $0.0079 \ Hz$, are nearly Gaussian and centered around zero. Outliers make up a 0.012 fraction of the data and their occurrence varies strongly over different time intervals. This is likely due to the increased difficulty in modeling the spacecraft's antenna and center of mass position \citep{Cascioli2021}. 

\begin{figure*}[h]
    \centering
    \includegraphics[width=\textwidth]{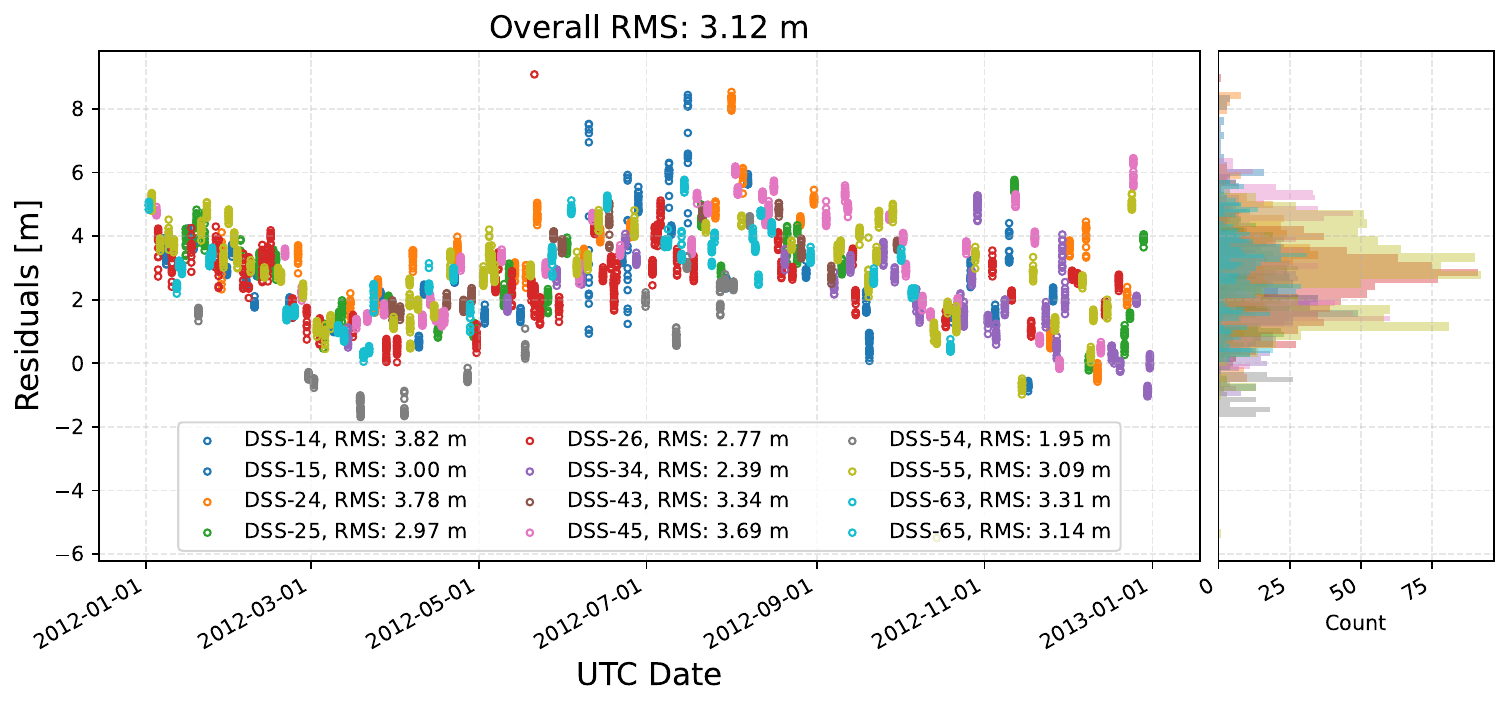}
    \caption{MRO sequential range residuals over a year. Different link end receivers have different colors.}
    \label{fig:mroRangeSpice}
\end{figure*}

\autoref{fig:mroRangeSpice} shows the residuals of the sequential range for the MRO over the same period, with outliers more than 10 meters removed. For this observable, the media effects are significantly stronger than for Doppler. Despite implementing a solar corona correction defined by \cite{Verma2012} with a single power law for the electron density, a clear signature from the solar opposition (when the Sun-Earth-probe (SEP) angle is near $180^\circ$) is visible around March 15th 2012. Although there is no other clear systematic signature, the residuals have a nonzero mean of a few meters (RMS$\approx$3 m). This is roughly consistent with findings for other Mars orbiters; for instance, estimated range biases for the Mars Global Surveyor (MGS) and Mars Odyssey missions are of the same order of magnitude (see Fig. 20 of \cite{Konopliv2006}).

Despite the minor discrepancies noted in the Doppler residuals and the non-zero-mean range residuals, the overall results are consistent with previous analyses for these missions. This validates Tudat models, and verifies that they have reached a level of maturity that is compatible with the current needs of the field. It also indicates that further improvements in the details of the models would allow for further refinement of the results.

\subsubsection{Open-Loop Doppler}
The PRIDE experiment provides an extension to typical radio tracking by employing a network of radio astronomical telescopes that detect the spacecraft downlink signal. From these detections, open-loop Doppler data at each station, as well as VLBI data of the spacecraft can be acquired \cite{Gurvits2023}. Due to the differences in the manner in which the data is acquired , the open-loop observable is modelled slightly differently compared to typical DSN/ESTRACK closed-loop data \cite{BocanegraBahamonEtAl2017}.

\begin{figure*}[h!]
        \centering
        \includegraphics[width=\textwidth]{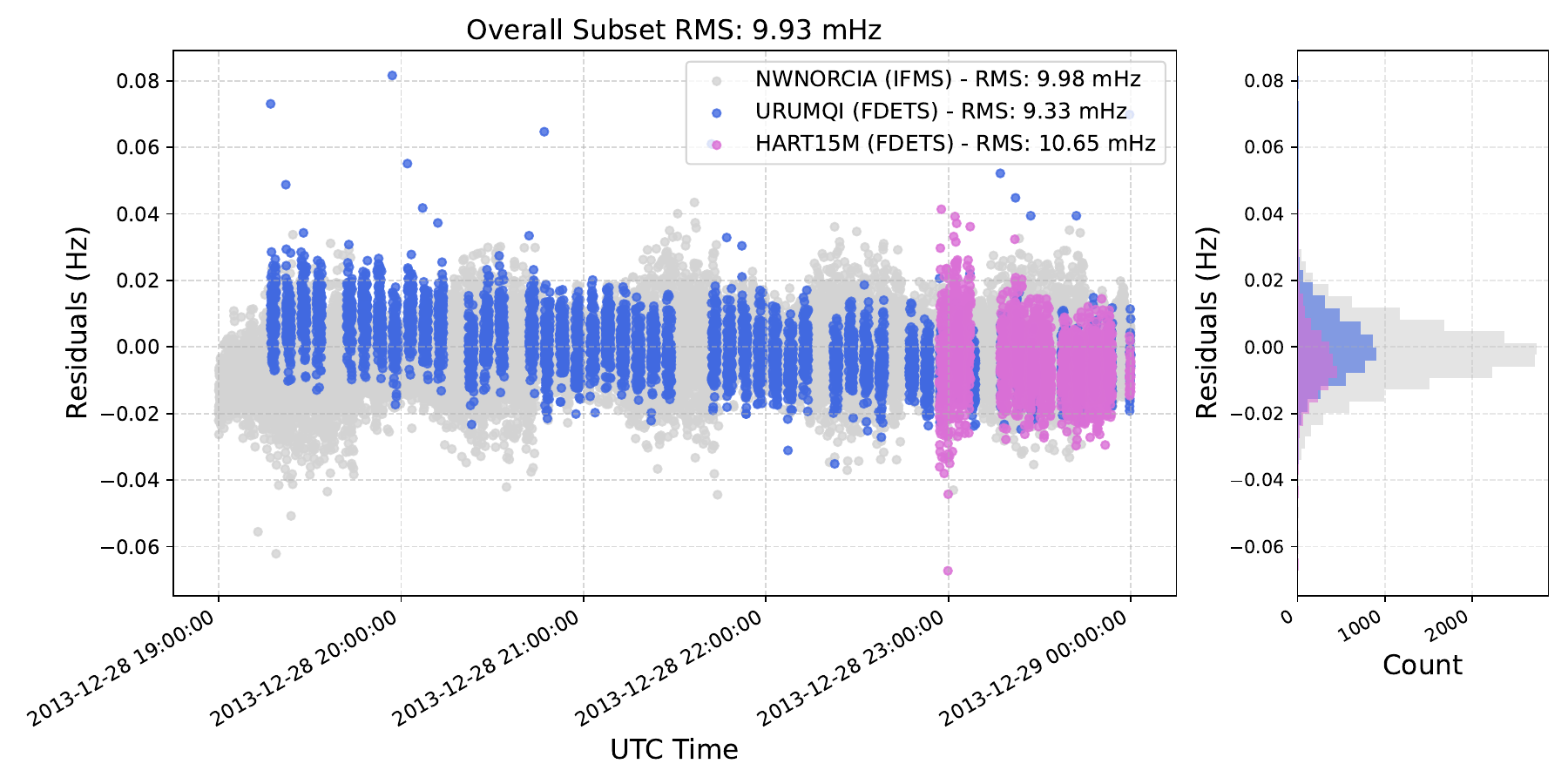}
        \caption{GR035 residuals for New Norcia, Urumqi, and Hart-15m stations on 28/12/2013. Left: Residuals as a function of time. Right: Residuals Distribution. The RMS is expressed in Hertz.}
    \label{fig:NN_HT_ON_Residuals}
\end{figure*}

To present Tudat's open-loop Doppler modeling capabilities, we show residuals obtained for Mars Express during the GR035 experiment (\citep{BocanegraBahamonEtAl2017}, \citep{Duev2016}), corresponding to the spacecraft's closest-ever flyby of Mars's moon Phobos. Frequency detections were recorded by PRIDE VLBI stations around the world during a span period going from 19:17:05 UTC on December 28th to 18:22:28 on December 29th 2013. During this time, PRIDE receivers shadowed the DSN and Estrack uplinks, and observations were taken with a cadence of 1 second. The corresponding data set is stored at JIVE in the form of Fdets files \citep{MoleraPhD}. Similarly to what was already presented in the previous section, settings for the Earth's rotation model and planetary ephemerides, together with the relevant light-time corrections were used. Simulating open-loop observables for Hart-15m and Urumqi and taking the difference between those and the Doppler detections, yielded an RMS for the residuals of about $0.01$ Hz for a compression time of 1 s, consistent with what was also presented in \cite{BocanegraBahamonEtAl2017}. The remaining trend is possibly associated with imperfect quality of the MEX reference trajectory. Comparing open-loop Doppler with closed-loop ones taken over the same time span and provided by the New Norcia antenna, we see that the two yield similar results, thereby confirming consistency in our model implementation for both techniques (see Figure \ref{fig:NN_HT_ON_Residuals}).

\subsection{Astrometric Observables}
\label{sec:data_processing:astrometry}
Astrometric observables are typically expressed in terms of Right Ascension and Declination, or alternatively through topocentric coordinates such as Azimuth and Elevation. These measurements arise from passive, optical detections and represent the foundational data used to determine the position and motion of interstellar and near-Earth objects. As such, they are particularly well suited for planetary defense initiatives and space debris detection. To demonstrate how astrometric data is ingested and processed within Tudat, we examine the case of Eros, a near-Earth asteroid approximately 34 kilometers in diameter. First detected in the 19th century, Eros has since been extensively studied, with over 9000 astrometric observations publicly available through the Minor Planet Center (MPC). Notably, measurements of Eros's parallax were used in the early 20th century to help refine the value of the Astronomical Unit (AU) \cite{WILSON1904}. In this subsection, our focus lies in the data preparation required for orbit estimation. The resulting estimation outcomes are presented separately in Section \ref{sec:applications:asteroids}. 

\begin{figure}[htbp]
        \centering
        \includegraphics[width=\columnwidth]{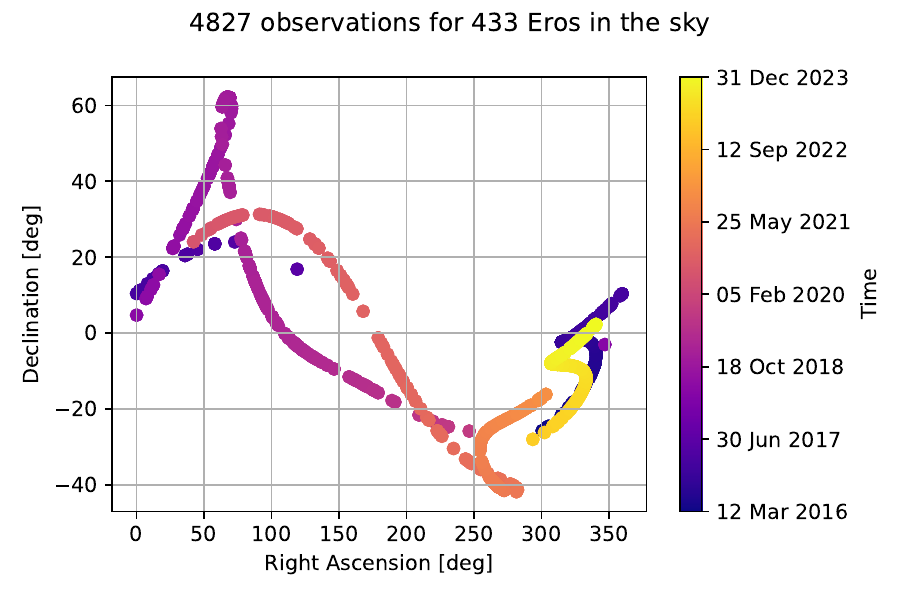}
        \caption{Astrometric observations (Ra and Dec) used in the state estimation of Eros.}
        \label{fig:estimation:planetary-defense:observations-sky-overview}
\end{figure}

Using Tudat's MPC interface (see \autoref{sec:data-ingestion:mpc-sbdb-horizons}) observations of Eros are queried and parsed in the time span from 1st January 2015 to 1st January 2024. The temporal evolution of all retrieved measurements is shown in Figure \ref{fig:estimation:planetary-defense:observations-sky-overview}.

\begin{figure}[t]
    \centering
    \begin{subfigure}[b]{1\columnwidth}
        \includegraphics[width=\linewidth]{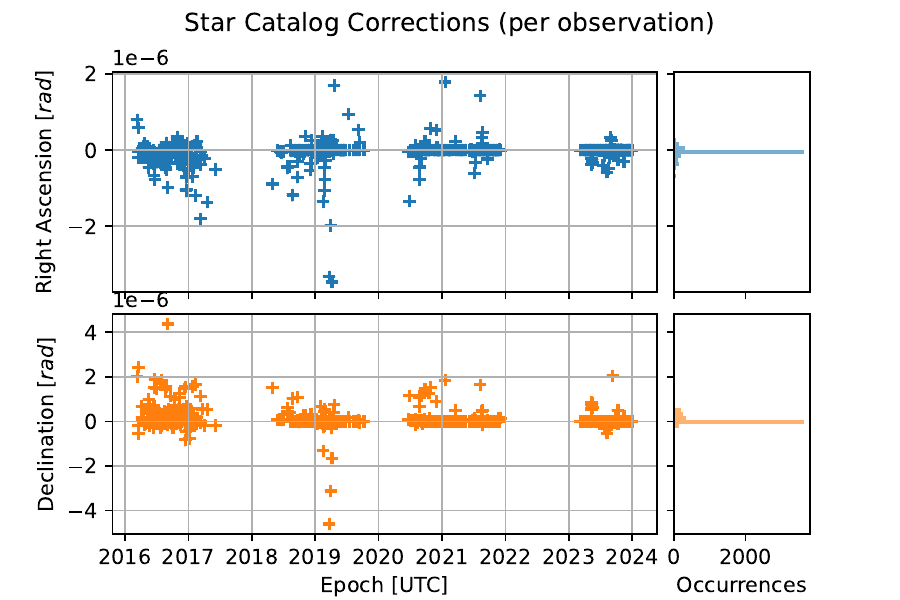}
        \caption{Star catalog corrections applied to the observations, following the method presented by \cite{Eggl2020}.}
        \label{fig:estimation:planetary-defense:star-catalog-corrections}
    \end{subfigure}
    \hfill
    \begin{subfigure}[b]{1\columnwidth}
        \includegraphics[width=\linewidth]{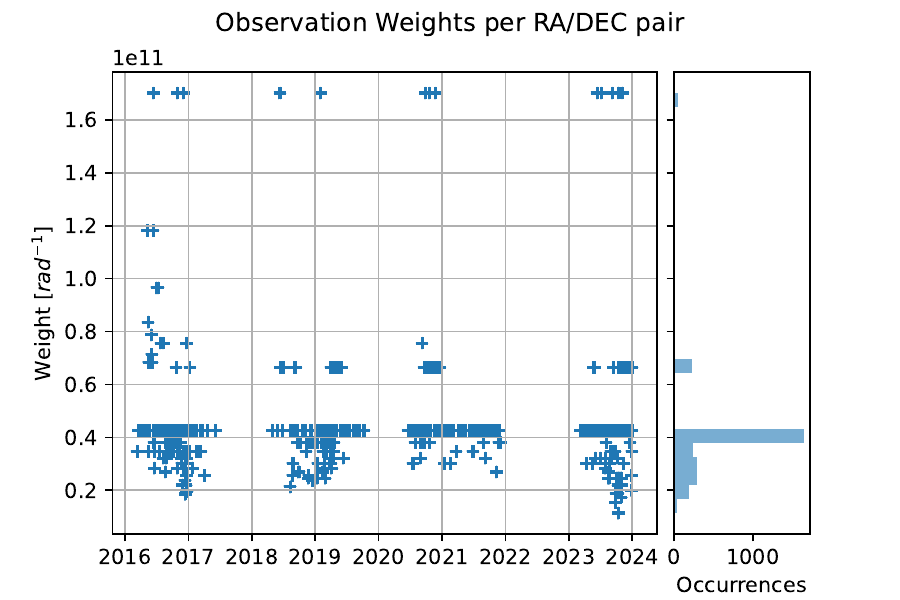}
        \caption{Observation weights assigned to observations, following the methodology presented by \cite{Veres2017}.}
        \label{fig:estimation:planetary-defense:observation-weights}
    \end{subfigure}
    \caption{Corrections and weights applied to asteroid observations. Panel (a): star catalog debiasing; Panel (b): telescope-specific weighting.}
    \label{fig:estimation:planetary-defense:corrections-weights}
\end{figure}

Before astrometric observations are ingested into the estimation algorithm, they may first be corrected, depending on the star catalog to which they are referenced. Older observations can be adjusted by incorporating improved knowledge from up-to-date star catalogs, thereby removing systematic biases. Corrections applied to right ascension and declination are shown in \ref{fig:estimation:planetary-defense:star-catalog-corrections}.
As it can be seen in \ref{fig:estimation:planetary-defense:star-catalog-corrections}, recent observations tend to have smaller corrections applied, suggesting that these observations were reduced with more recent star catalogs, requiring smaller corrections (if any).
Once corrected for star catalog biases, the observations are converted to Tudat-native data structures, which link the observation to the corresponding telescope.
Lastly, when converting the observations to Tudat-native data structure, weights can be optionally applied to each observation, depending on the observatory. Tudat uses the method published by \cite{Veres2017} to weigh each observation according to the observatory it has been recorded from, how many observations during the night have been recorded and, if applicable, the epoch at which it has been recorded. The weights assigned per observation are shown in \ref{fig:estimation:planetary-defense:observation-weights}.


\section{Tudat's Orbit Estimation Capabilities}
\label{sec:EstimationCapabilities}
In this section, we showcase Tudat's orbit estimation capabilities for both radiometric and optical data presented in \autoref{sec:radiometric_observables}. In \autoref{sec:radiometric_estimation} we present the orbit determination results for the GRAIL and MRO missions, demonstrating the effectiveness of Tudat in supporting missions both in the Lunar environment and in the deeper solar system.  In \autoref{sec:Astrometric_Estimation} we further extend this analysis to include astrometric data for the orbit determination of passive bodies. Specifically, the estimation of the orbit of the asteroid Eros and the re-entry trajectory of the Kosmos 482 descent craft will be discussed.

\subsection{Radiometric Estimation}
\label{sec:radiometric_estimation}

Using the Closed-Loop Doppler datasets introduced in Section \ref{sec:radiometric_observables}, we present orbit estimation results for representative arcs of both the GRAIL and MRO spacecraft. GRAIL is a NASA lunar science mission that used high-precision gravitational field mapping of the Moon to investigate its interior structure \cite{Asmar2014}. Consequently, orbit estimation of GRAIL within Tudat offers a meaningful test for the dynamical model in the Earth-Moon environment. In contrast, MRO, which operates around Mars, serves as a validation case for Tudat’s capabilities in deep-space dynamics.

\subsubsection{Lunar Mission - GRAIL}
\label{subsec:grailestimation}
The GRAIL reference trajectory in the Spice kernel was estimated using a combination of Doppler and data from the inter-satellite K-band ranging link (which provides much more accurate tracking than the DSN Doppler data). As a result, the reference orbit's accuracy is expected to be well below that of our analysis, and the difference of our post-fit orbit is a robust measure for our Doppler-only estimation error.

Here, we show the results of the orbit estimation of the GRAIL-A spacecraft over a one week period. The dynamical environment is well characterized in the literature thanks to the GRAIL and LLR datasets. Our model includes point-mass gravity from the Sun and all planets, a high-fidelity spherical harmonic gravity field as associated body fixed frame for the Moon, and accelerations due to radiation pressure following \cite{LEMOINE2013}. For solar radiation pressure, we use a spacecraft macro-model that defines the size and orientation of each panel \cite{Fahnestock2012}. We also include comprehensive models for lunar radiation, accounting for both albedo and infrared emissions \cite{FLOBERGHAGEN1999,LEMOINE2013}. For each arc, we estimate the spacecraft's initial state and a total of four scaling factors for radiation pressure (two for solar radiation, two for moon radiation, one parallel with the vector to the source body, one perpendicular to it), as well as the amplitude of maneuver, when applicable. For the analyzed time interval, maneuvers are present and thus estimated in the first and last arcs at epochs 2012-04-06 13:18:47 and 2012-04-12 07:28:37, respectively.

\begin{figure*}[h!]
\centering
         \captionsetup{skip=1pt}
    \begin{subfigure}[b]{0.85\textwidth}
        \includegraphics[width=\textwidth]{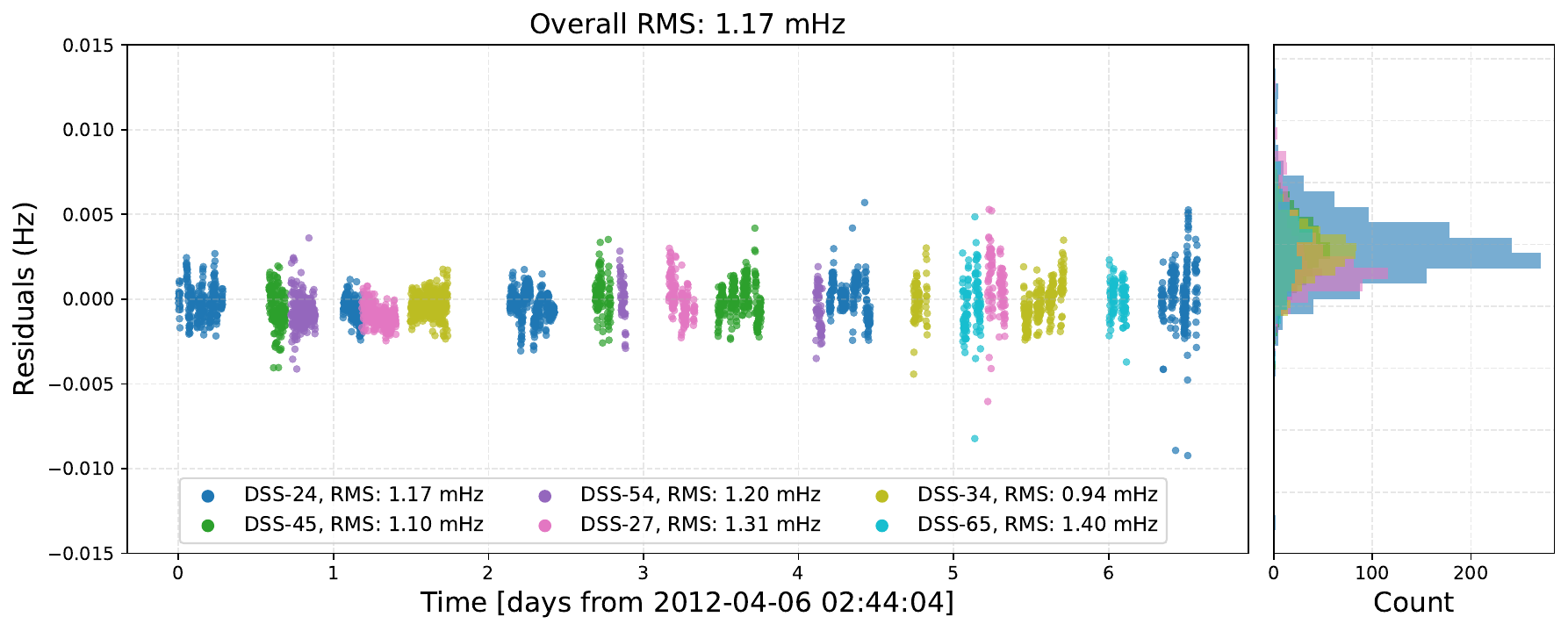}
        \caption{GRAIL averaged Doppler post-fit residuals w.r.t SPICE. The RMS residuals are expressed in millihertz.}
        \label{fig:grailPostFitResiduals}
    \end{subfigure}
    \vfill
    \begin{subfigure}[b]{0.85\textwidth}
        \includegraphics[width=\textwidth]{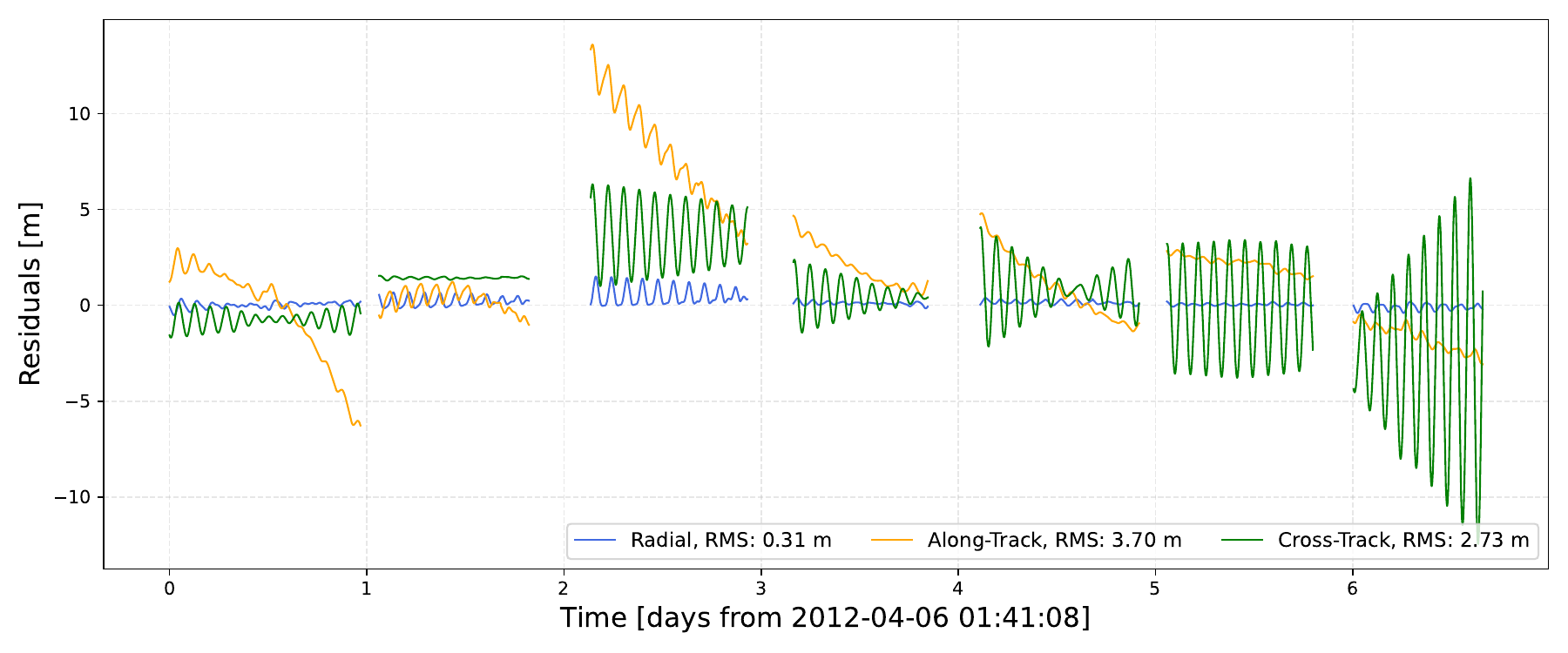}
        \caption{GRAIL post-fit states (orbit) residuals wrt to SPICE.}
        \label{fig:grailPostFitStates}
    \end{subfigure}    
    \caption{Resulting residuals for the GRAIL mission.}
    \label{fig:grailPostFit}
\end{figure*}



The results of our Doppler data analysis are shown in Fig. \ref{fig:grailPostFit}, displaying both the post-fit residuals of the Doppler data (Fig. \ref{fig:grailPostFitResiduals}, residuals greater than 0.01 Hz have been filtered out) and the difference between our estimated orbit and the GSFC reference orbit (Fig. \ref{fig:grailPostFitStates}). The difference in radial position is on average at the level of 10 cm, whereas long- and cross-track errors are at the level of 1 m. Notably, the arc on 2012-04-08 shows a significant deviation from this average, which requires further analysis. The post-fit Doppler residuals are around 1 mHz RMS (which matches the expected level of the data noise). For reference, a 0.1 mm/s range-rate error would result in a Doppler shift of about 1.5 mHz. The post-fit orbit error and the level of the observation residuals match the expected quality of the orbits and the data well, validating the suitability of Tudat as an open-source orbit estimation toolbox for lunar missions.

\subsubsection{Deep Space Mission - MRO}
\label{subsec:mroestimation}
\begin{figure*}[h!]
    \centering
    \begin{subfigure}{\textwidth}
        \centering
        \includegraphics[width=\linewidth]{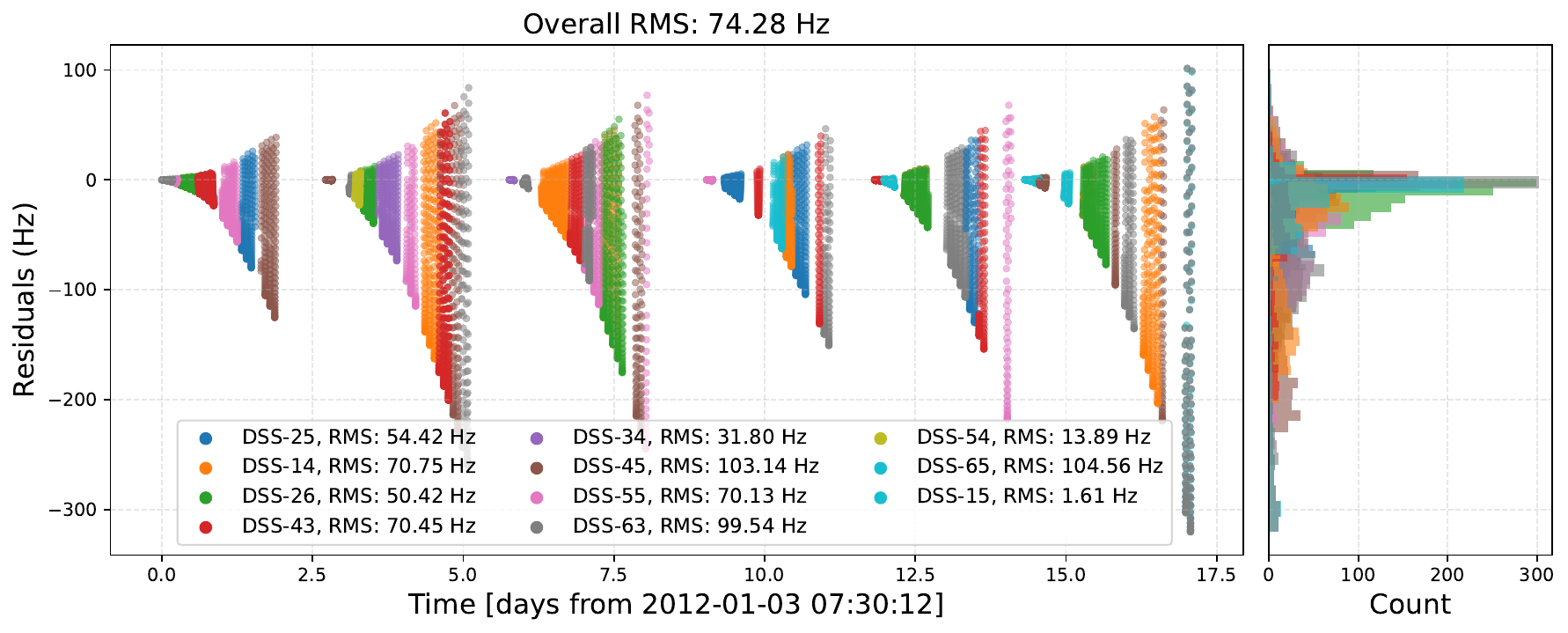}
        \caption{}
        \label{fig:mroprefitres}
    \end{subfigure}
    \begin{subfigure}{\textwidth}
        \centering
        \includegraphics[width=\linewidth]{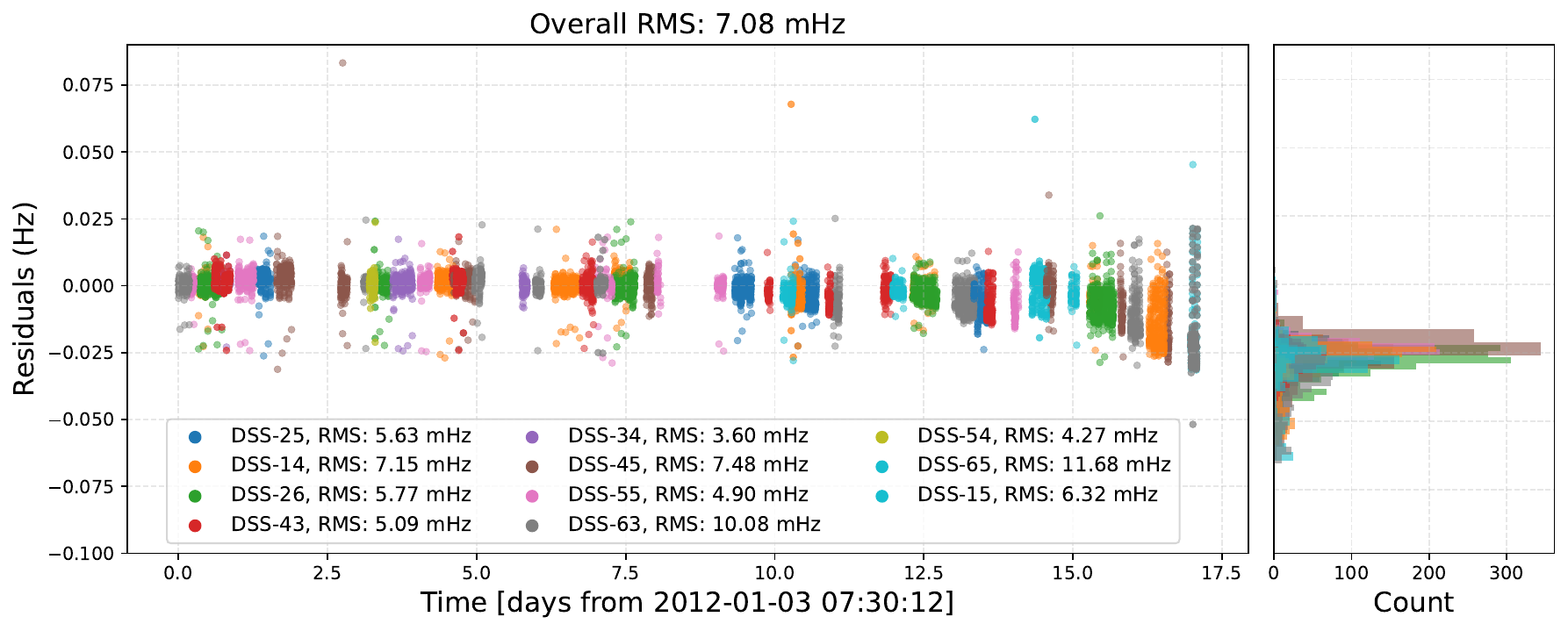}
        \caption{}
        \label{fig:mropostfitres}
    \end{subfigure}
    \caption{MRO Averaged Doppler pre-fit (a) MRO  post-fit residuals (b) of the orbit determination process.}
    \label{fig:mro_residuals_od}
\end{figure*}

\begin{figure*}[h!]
    \centering{
        \includegraphics[width=\textwidth]{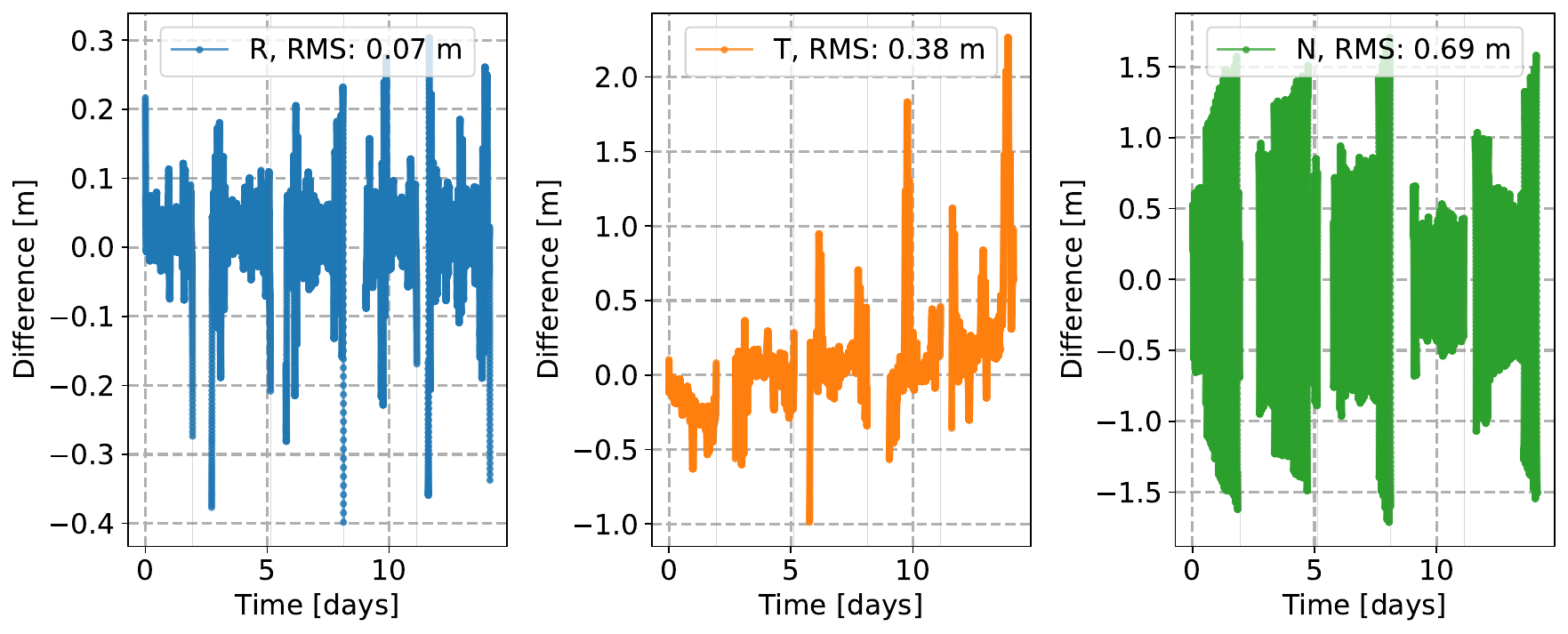}
    }
    \caption{MRO post-fit states (orbit) residuals wrt to SPICE.}
    \label{fig:mro_orbit_pre_post_fit}
\end{figure*}

Here, we present the results of the orbit determination of the MRO spacecraft over a period from 2012-01-01 to 2012-01-19. The approach follows proven POD strategies from literature \cite{Highsmith2008,Johnston2016,Konopliv2016,Genova2016}, using two- and three-way DSN Doppler data data and ancillary files from the official MRO Gravity/Radio Science PDS archive. MRO is subject to significant non-gravitational forces, primarily atmospheric drag resulting from its low-altitude orbit, and solar radiation pressure. Accurate modeling of these forces is the primary challenge in achieving high-precision trajectory solutions \cite{Mazarico2008,Mazarico2009}. Our dynamical model includes the JPL JGMRO120D gravity field complete to degree and order 120, solid body tides, and the associated Mars rotation model \cite{Konopliv2016}. For non-conservative forces, we model the Martian atmosphere using the DTM-Mars model \cite{Bruinsma2002} and employ a detailed spacecraft macro-model consisting of the bus, high-gain antenna, and solar panels. The spacecraft orientation and inter-plate self-shadowing effects are modeled using reconstructed attitude C-kernel files from the navigation team.
We process data over 7 contiguous arcs of approximately two days each, with arc lengths chosen to avoid angular momentum desaturation maneuvers. For each arc, we estimate the spacecraft's initial state and scaling factors for solar radiation pressure and atmospheric drag to account for force model uncertainties. Additionally, empirical accelerations (constant, sine, and cosine components) are estimated in the along-track and cross-track directions to absorb residual modeling errors.
As shown in \autoref{fig:mropostfitres}, the pre-fit Doppler residuals, generated using the nominal force models, have RMS of 73.6 Hz and exhibit a distinct diverging signature. The orbit determination process reduces these residuals by four orders of magnitude to an RMS of $7.77 \times 10^{-3}$ Hz. This result is comparable to the residuals relative to the navigation team's official orbital solutions ($7.90\times 10^{-3}$ Hz as shown in \autoref{fig:mroDopplerSpice}). Consequently, our reconstructed orbit is in good agreement with the solution of the navigation team, with the resulting position differences contained within $\pm0.4 \ m$± radially and a few meters in the components along and across the track, as shown in \autoref{fig:mro_orbit_pre_post_fit}. However, the residuals in the final two arcs exhibit a diverging signature similar to that of the pre-fit data, warranting further investigation.

\subsection{Astrometric Estimation}
\label{sec:Astrometric_Estimation}
Similarly to the estimation from radiometric Doppler data (\autoref{sec:radiometric_estimation}), this Section showcases Tudat's astrometric estimation capabilities. Specifically, we present results from the orbit determination of asteroids and SSA applications.

\subsubsection{Applications to Asteroids}\label{sec:applications:asteroids}

We use the same MPC Eros dataset that we presented in \autoref{sec:data_processing:astrometry}, spanning the period between January 1, 2015 and January 1, 2024. 
The dynamical model includes the gravitational point-mass accelerations of the Sun, all planets, the major moons of Jupiter and Saturn, as well as relativistic corrections for the Sun's Schwarzschild effect. We have not performed a detailed analysis of the perturbations exerted on Eros by other asteroids. 

\begin{figure*}[!h]
     \captionsetup{skip=1pt}
        \includegraphics[width=\textwidth]{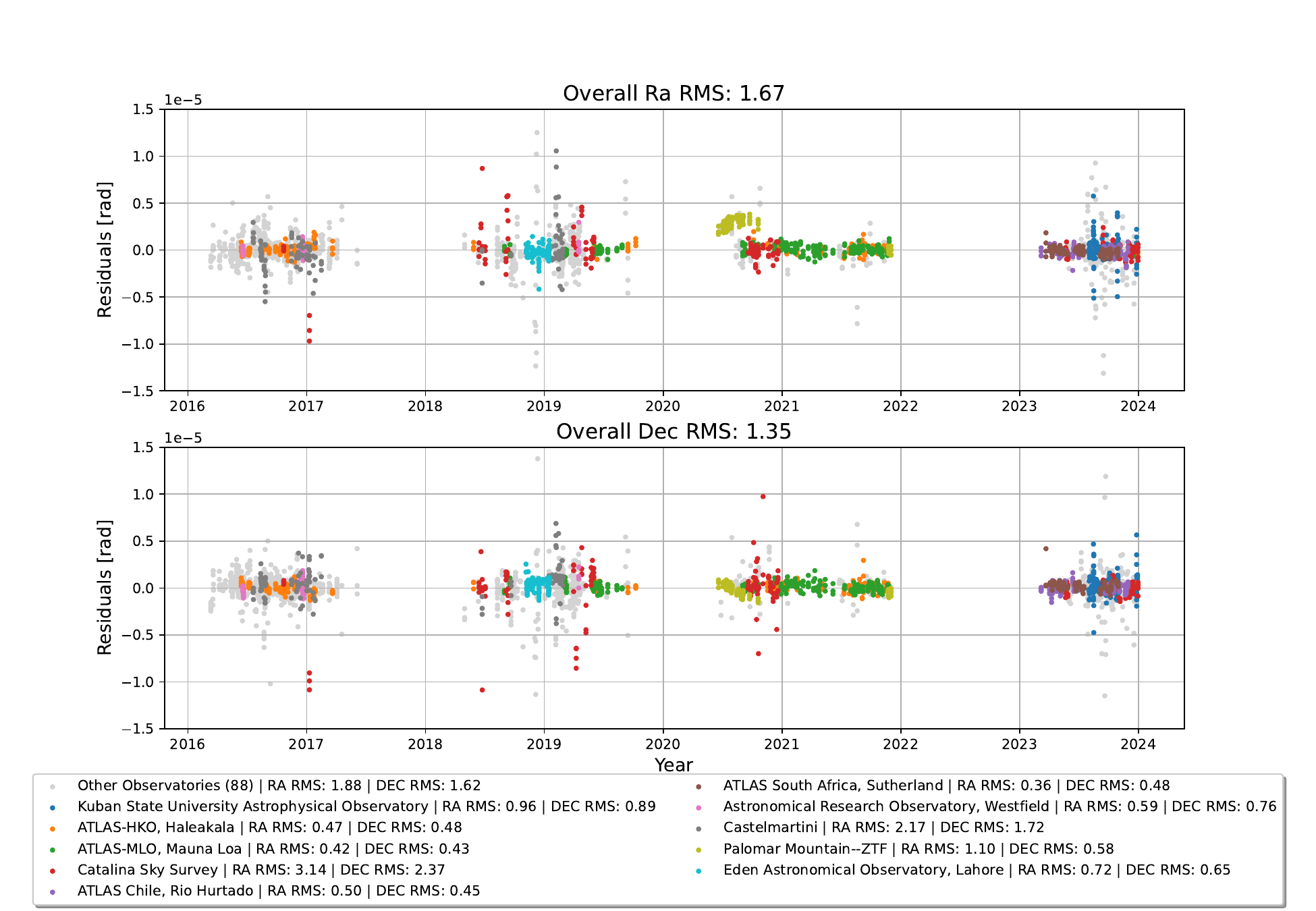}
        \caption{Eros post-fit residuals w.r.t JPL Horizons. (Top) Right Ascension. (Bottom)
        Declination. The RMS of the residuals is expressed in arcseconds.}
        \label{fig:ErosPostfitResiduals}
\end{figure*}

We only estimate Eros' translational state at the reference epoch, with no additional (consider) parameters. To emphasize the influence of observation preprocessing on the final orbit estimates, we vary the preprocessing configuration across multiple estimations. Specifically, we consider the following scenarios: (i): no star catalog correction, no weighting applied; (ii): star catalog correction applied, without weights; (iii): star catalog correction and weighting applied. The corresponding results are shown in Fig.~\ref{fig:estimation:planetary-defense:rsw-error-jpl}, where the difference of the estimated orbit compared to the JPL Horizons solution is plotted. For completeness, we also showcase the computed astrometric post-fit residuals, see Figure \ref{fig:ErosPostfitResiduals}

As can be seen from Fig. \ref{fig:estimation:planetary-defense:rsw-error-jpl}, applying the star catalog correction (case (ii) and case (iii)) reduces the difference w.r.t. to JPL Horizons, although to a modest degree.
Applying weights to each observation does not consistently improve the errors compared to the JPL solution, and in some cases leads to larger errors compared to the unweighted case. This warrants further investigation, to be carried out in the future.
In the following, case (iii) (i.e. including star catalog corrections and observation weighting) will be analyzed in more detail.
The corresponding formal errors of the solution obtained with Tudat are shown in the bottom plot of Fig.~\ref{fig:estimation:planetary-defense:rsw-error-jpl}.
The difference between our solution and the JPL Horizons solution mostly falls within the 3-sigma uncertainty of our model, but exceeds the 3-sigma uncertainty of the JPL solution.\footnote{When considering formal errors, one should always keep in mind that they are typically an optimistic measure of the true error.} 
Since we expect the JPL solution to be substantially better than ours (see below), we can assume the difference between the two solutions to be a measure of the error of our solution. Since this difference also lies mostly within the 3-sigma error bounds of our solution, we conclude that the solution statistics (our formal errors) provide a reasonable measure of its real error. Consequently, it indicates the reasonable suitability of our dynamical model.

Our orbital error falls within the range of several tens of kilometers, which corresponds to an angular separation between our result and the JPL result of about 20 milliarcseconds at a distance of 1 AU. In addition to a clear once-per-orbit trend in the orbit difference, we see a clear secular trend in the S-component (i.e., the along-track component of the error; top of Fig.~\ref{fig:estimation:planetary-defense:rsw-error-jpl}), suggesting that there is either a long-term periodic or secular trend in the error of our solution.
However, considering the limitations we have chosen for our setup, the observed behaviour is not unexpected.
Our estimation results presented here only include data from an interval of eight years ($\approx$ 4.5 orbital periods of Eros), while the solution produced by JPL includes observational data starting from 1893.
In addition, the solution obtained using Tudat does not include any radar observations, which would help to better constrain the orbit, as they provide information along the telescope-asteroid line of sight. Also, space-based observations (for instance, from TESS) are not included. Lastly, no sophisticated outlier rejection or weighing scheme based on the data itself has been incorporated in the presented analysis. 
As we also discuss in Section \ref{sec:future_developments}, the functionality for processing astrometric data of small solar system bodies in Tudat will be extended and applied much more deeply in the future.

\subsubsection{Applications to Space Situational Awareness}
\begin{figure*}[!h]
\hspace*{-1cm}
\includegraphics[width=1.1\textwidth]{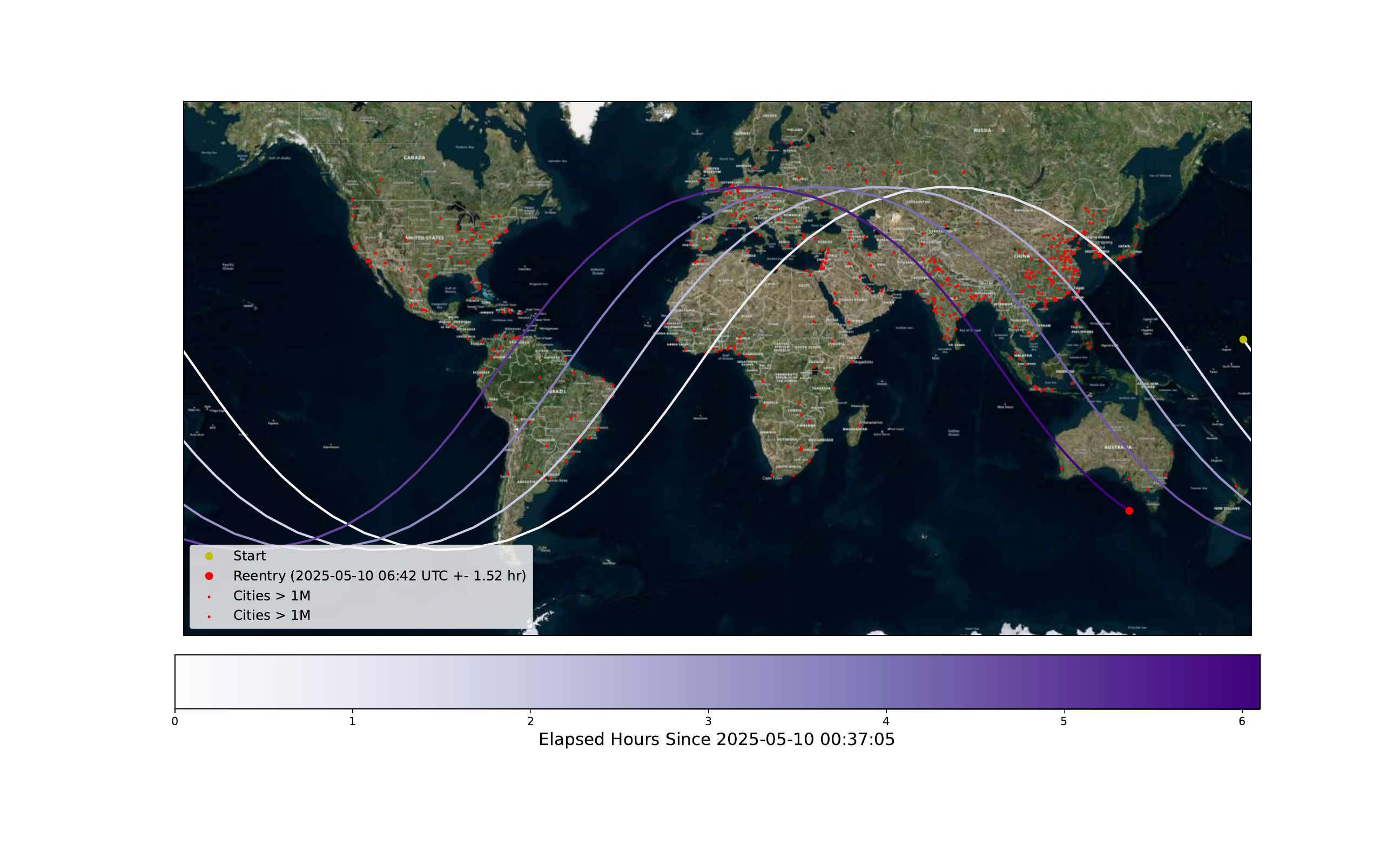}
\captionsetup{skip=1pt}
\caption{Kosmos 482 Ground Track on 2025-05-10. World cities with a population greater than 1 million are highlighted.}
\label{fig:kosmos_482}
\end{figure*}

\begin{figure*}[h]
    \centering
    \includegraphics[width=\textwidth]{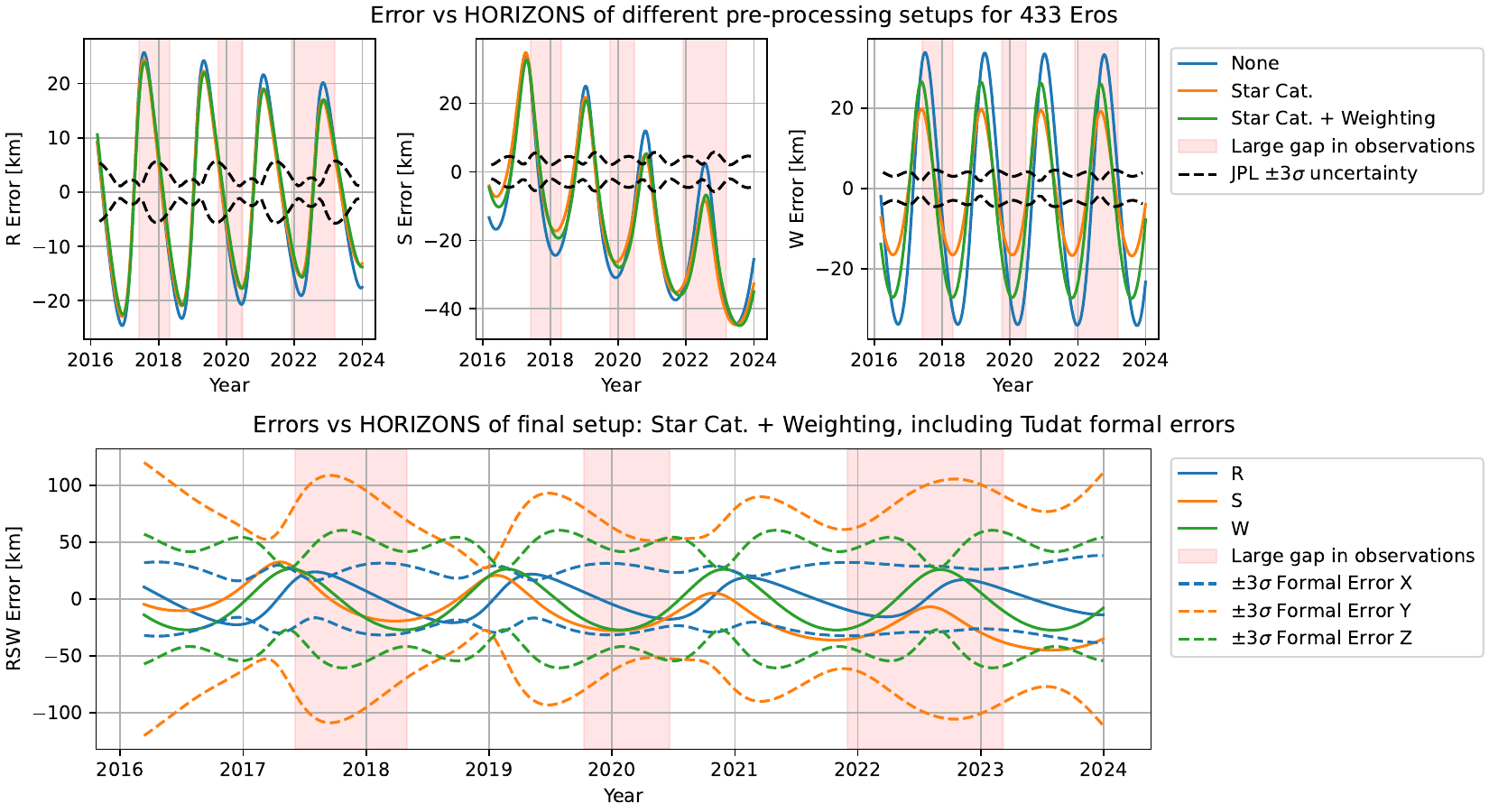}
    \caption{Comparison of orbital estimation errors with respect to JPL Horizons. The upper plots compare the different pre-processing techniques ("Star Cat." corresponding to debiasing of observations depending on reference star catalog, "weighting" corresponding to weighting of the observations based on the telescope.), the bottom plots shows the best-performing setup from the top plots, including the corresponding formal errors of the Tudat solution.}
    \label{fig:estimation:planetary-defense:rsw-error-jpl}
\end{figure*}

The uncontrolled reentry of a unique object, the Kosmos 482 Descent Craft (COSPAR 1972-023E, SSC cat. nr. 6073) on 10 May 2025, was used as a test case in exploring the use of Tudat for long-term geocentered orbital evolution simulation and reentry forecasting. The Kosmos 482 Descent Craft was the lander module of a failed 1972 Soviet Venera mission that got stuck in an initially 9800 x 210 km, 51.7 degree inclined Highly Elliptical Earth Orbit. The lander separated from the main bus in June 1972 \cite{Langbroek2022}. The lander was semi-spherical in shape, meaning that the drag area for this object was at worst only modestly variable, making the use of a cannonball model for aerodynamic and radiation pressure forces a reasonable assumption. We created a reentry forecast model in Tudat employing the NRLMSISE00 model atmosphere \cite{Picone2002}. Our final reentry forecast, based on a last available orbit update with an epoch six hours before our projected reentry moment, was 10 May 2025 at 6:42 UTC (± 1.52 hours). Figure \autoref{fig:kosmos_482} shows Tudat's computed Kosmos 482 ground track for that day. This is close to the reentry model results of various other organizations, and within the window 6:04 - 7:32 UTC established by first a positive and then a negative radar observation of the object from Germany, as reported by ESA. Unfortunately, no final high-accuracy Time of Impact Prediction (TIP) was issued by CSpOC this time (and the last TIP was clearly erroneous, given the German radar observation outside the reentry interval of that TIP).

Our analysis shows that, for near-term entry predictions, the models and capabilities in Tudat provide valuable tools. Combined with our orbit estimation framework, this provides a powerful open-source tool for near-Earth SSA.

\section{Future Developments}
\label{sec:future_developments}

In the previous sections we have demonstrated that Tudat has capabilities for high-fidelity orbit propagation, estimation and prediction over a broad range of regimes and objects, as well as for various different data sets. The developments to allow Tudat to process real data, leading to results presented, has largely been implemented by a small team in the period 2022-2025, building on about 10 years of development for the overall framework in the context of (mostly) simulations. 

Now that the core functionality for using real tracking data for orbit estimation is available and validated in Tudat, we will focus on:
\begin{itemize}
\item Refining, enhancing and diversifying existing models, including performing more rigorous validation of our models and tools against other software;
\item Improving the software architecture and interfaces of Tudat to allow it to be more easily used for large-scale analyses;
\item Using the functionality in Tudat for a diverse range of applications, focusing on SSA and planetary science;
\item Further improving the documentation and accessibility of Tudat by making it fully comprehensive (covering all available functionality) and providing better training material and tutorials (including videos).
\end{itemize}

The core modelling of Tudat in terms of both observations and dynamics has been thoroughly tested and validated by ~300 unit tests, as well as various \textit{ad hoc} validations done during research projects (such as \cite{Fayolle2023}). For planetary tracking data, Tudat is currently being used in projects for the analysis of Cassini tracking data \cite{Hener2025b}, Rosetta tracking data \cite{Reichel2025}, and Martian orbiter tracking data \cite{Alkahal2024}. For small solar system bodies, we plan to start exploratory projects to leverage Tudat for high-fidelity estimation of Near-Earth objects \citep{Verdoeskleijn2025}, and non-conservative force modelling of both asteroids and comets. For near-Earth SSA applications, work is currently undergoing to set up the analysis framework to process astrometric data of Earth-orbiting spacecraft and debris, from both an all-sky camera and a narrow-field tracking camera, to provide daily updates to the orbits of thousands of objects per night. For the latter, we will (among others) develop automatic interfaces to the \href{https://www.space-track.org/auth/login}{Space-Track.org} database and \href{https://discosweb.esoc.esa.int/}{ESA's DISCOS} database, as well as implement initial orbit determination algorithms to allow orbit refinement for objects that are not identified. 

For what concerns dynamical modelling, we have recently extended our capabilities in terms of non-conservative force modelling to include self-shadowing for both radiation pressure and aerodynamic forces \citep{balmino2018}. In addition, we have unified our framework for computing these accelerations for both types from panel methods and have improved and streamlined our interfaces to use these models. For aerodynamic modelling, these new interfaces permit the implementation of panelled force computations over multiple regimes, facilitating the propagation of an object from rarefied flow into hypersonic flow with an automatic bridging of the coefficients based on different gas-surface interaction models (GSIMs) \citep{Sinpetru2022}. The results presented for MRO above used a preliminary version of this functionality. We will also work on linking different atmosphere models to Tudat (such as the Mars Climate Database \citep{Millour2022mcd}), and provide default surface albedo and infrared maps for additional bodies (currently only available for the Earth and Moon).

For the JUICE mission, Tudat will be used for the primary data processing tool of the PRIDE experiment \citep{Gurvits2023}, which provides Doppler and VLBI data. The dynamical model requirements to process all JUICE data are extremely stringent \citep{VanHoolst2024}, and will require a greater degree of consistency in concurrent orbital, rotational and tidal modelling for natural satellites. Such a framework is in development for Tudat \citep{Fayolle2025}, expanding the coupled orbital-rotational dynamics propagation for natural bodies that we have already used \citep{Martinez2025}. The application of Tudat for the JUICE mission will also require us to perform data fusion of different observations to a much greater degree than we have done thus far. Before the arrival of JUICE in the Jovian system, we will further develop the existing implementation and expertise based on past missions, such as Cassini. To further improve the data processing capabilities of Tudat, we will extend the file processing capabilities to include older radio tracking data in ATDF format, using the software presented by \cite{Verma2022}. Expanding on our MPC interface, we will also provide an interface to the Natural Satellite Data Center (NSDC, \cite{Arlot2009}) for applications to natural satellite ephemeris estimation, a first iteration of which was done in Tudat by \cite{Dahmani2024}. 

We plan to close the loop between the existing range models in Tudat and the radar data for small solar system bodies available from the MPC. With these expansions in place, the software will be able to incorporate a wider range of data sources and types, leading to enhanced estimation capabilities. Tudat is unique in providing the capability it does in a completely open-source manner. By further strengthening and diversifying its capabilities, the project opens up the field to a much broader community. The research projects that have recently kicked off using the new functionality will all follow a similar publication paradigm: along with publications, there will be a link to a repository with all relevant code, where our published results can be reproduced by anyone with minimal effort. This can be advantageous to those trying to set up a similar application, or those trying to improve upon our work. But it is also very important for researchers who use our orbit estimation results. By providing full openness in the results and models, projects doing (for instance) interior determination from gravity field coefficients, Love numbers, rotational parameters, \textit{etc.} can perform their own (re)analyses with different settings and, perhaps more importantly, carry out an independent analysis of the uncertainty bounds of the solutions, to ensure a robust mapping from data noise to geodetic parameter errors to interior property uncertainties. 
Finally, we aim to provide an important incentive to the broader community to follow suit and adopt more openness in the use of flight-dynamics software for scientific applications. 

\section*{Acknowledgements}
We gratefully acknowledge the valuable
feedback provided by the many students who have engaged with Tudat throughout the 'Numerical Astrodynamics' and 'Propagation and Optimization in Astrodynamics' courses, as well as during their thesis work within the M.Sc. Spaceflight Dynamics program at the Aerospace Engineering faculty of Delft University of Technology.

\bibliography{biblio}

\end{document}